\documentclass[prd,aps, preprintnumbers, showpacs, nofootinbib,notitlepage,twocolumn]{revtex4-2}
\usepackage[normalem]{ulem}
\usepackage{epsfig}
\usepackage{amsfonts}
\usepackage{amsmath}
\usepackage{slashed}
\usepackage{graphicx}
\usepackage{color}
\usepackage{mathtools}
\usepackage{simpler-wick}
\allowdisplaybreaks[4]

\newcommand{\vect}{\boldsymbol}

\begin{document}
	
	\preprint{JLAB-THY-22-3728}
	
	\title{Counting linearly polarized gluons with lattice QCD}

	\author{Shuai Zhao}
	\email{zhaos@tju.edu.cn}
	\affiliation{Department of Physics, Tianjin University, Tianjin 300350, China}
	\affiliation{Department of Physics, Old Dominion University, Norfolk, VA 23529, USA}
	\affiliation{Theory Center, Thomas Jefferson National Accelerator Facility, Newport News, VA 23606, USA}
	
	\begin{abstract} 

We outline an approach to calculate the transverse-momentum-dependent distribution  of linearly polarized gluons inside an unpolarized hadron on the lattice with the help of large momentum effective theory. To achieve this purpose, we propose calculating a  Euclidean version of the degree of polarization for a fast-moving hadron on the lattice, which is ultraviolet finite, and no soft function subtraction is needed.  
It indicates a practical way to explore the distribution of the linearly polarized gluons in a proton and the linearly polarized gluon effects in hadron collisions on the lattice. 

	\end{abstract}

	\date{\today}

	\maketitle

%\section{Introduction}

It has been widely accepted that hadrons are constructed by quarks and gluons. Due to the nonperturbative nature of strong interaction, it is hard to explore how those building blocks combine a hadron. In high-energy processes, the parton information is encoded in the parton distribution functions (PDFs), which are one-dimensional distribution functions that describe the longitudinal momentum distribution of partons. If the partons are not collinear to the mother hadron but carry transverse momenta,  then the parton structure should be described by the transverse momentum dependent distributions (TMDs)~\cite{Collins:2011zzd}. The TMDs can describe much richer partonic structures of a hadron.

Gluon plays an important role in a proton.  Analog to a photon, a gluon can be unpolarized but also linearly polarized inside an unpolarized proton, if the transverse motion of gluon is considered~\cite{Mulders:2000sh}. The TMD for unpolarized gluon is denoted as $f_1^g (x,\vect{k}_T^2)$, while the TMD of linearly polarized gluon is denoted as $h_1^{\perp g}(x, \boldsymbol{k}_T^2)$, which can be regarded as the gluonic analog of the Boer-Mulders function~\cite{Boer:1997nt} for quark. The $T$-even function $h_1^{\perp g}$ describes how the $+1$ and $-1$ helicity gluon states are correlated in a hadron.

The linearly polarized gluon TMD has caused lots of attention recently. It has been pointed out that the linearly polarized gluons can modify the transverse spectrum of Higgs bosons and can be utilized to determine the parity of Higgs boson at the LHC~\cite{Boer:2011kf}.
%and heavy quarkonia~\cite{Boer:2012bt} which are produced in collisions of unpolarized protons.  
%Then, the essential question is how to determine it. 
In the past, 
the distribution of linearly polarized gluons inside an unpolarized hadron has been discussed in a model context in Refs.~\cite{Meissner:2007rx,Bacchetta:2020vty}, and
many approaches based on experimental observables are proposed to extract the gluon TMDs, e.g., heavy quark pair or dijet~\cite{Boer:2010zf,Pisano:2013cya,Efremov:2017iwh,Efremov:2018myn}, $\gamma\gamma$~\cite{Qiu:2011ai}, back-to-back quarkonium and photon productions~\cite{denDunnen:2014kjo},  quarkonium and dilepton associated productions~\cite{Lansberg:2017tlc}, single and double heavy quarkonium production at hadron colliders~\cite{Boer:2012bt,Ma:2012hh,Lansberg:2017dzg,Scarpa:2019fol}, etc. The effect of linearly polarized gluon TMD can be found in azimuthal asymmetries, and also in total cross sections.  Other previous studies  are devoted to the linearly polarized gluons at small-$x$ region~\cite{Dominguez:2010xd,Dominguez:2011br,Metz:2011wb,Marquet:2017xwy,Dumitru:2018kuw}. Although the gluon TMDs can be probed at the high-energy electron-ion colliders, e.g.,  EICs in the US and China,  however,
% at present,  we still have very little knowledge on the linearly polarized gluons in the hadron. However, 
the linearly polarized gluon TMD has never been extracted so far, either from the experiment or from lattice QCD.  

%There are some attempts at the lattice calculation of quark TMDs defined with straight-Wilson line operators; however, the Wilson %line in the TMDs appearing in the factorization theorems is always staple-shaped, which is different from the straight-line operators.

Measuring TMDs inside a hadron has been indicated possible due to the development of parton physics on the lattice in the past few years, which includes but is not limited to quasi-PDFs and Large Momentum Effective Theory (LaMET)~\cite{Ji:2013dva,Ji:2014gla}, pseudo-PDFs~\cite{Radyushkin:2017cyf,Radyushkin:2019mye}, lattice cross sections~\cite{Ma:2014jla}. These approaches have made significant progress on  PDFs, meson distribution amplitudes, generalized parton distributions, etc (see, e.g.,~\cite{Cichy:2018mum,Ji:2020ect} for recent reviews of LaMET). Especially, the TMDs defined with staple-shaped Wilson line operators have been considered recently within the framework of LaMET, see~\cite{Ji:2019sxk,Ji:2019ewn,Ebert:2019okf,LatticeParton:2020uhz,Ji:2021uvr,Zhang:2022xuw} and references therein.
The gluon-TMDs have also been considered very recently~\cite{Schindler:2022eva,Zhu:2022bja}.

In this work, we will show that evaluating the linearly polarized gluons inside an unpolarized hadron is feasible on the lattice with the help of LaMET. Our unambitious but practical idea is to calculate the ratio of $h_{1}^{\perp g}$ and the unpolarized gluon TMD $f_1^g$ in $(x,\vect{b}_T)$-space so that future lattice simulations can help to reveal the scale of the linearly polarized gluons.  To achieve this purpose, 
we define the Euclidean version of this ratio, which can be simulated on the lattice.  {In the large hadron momentum limit the degree of polarization can be recovered with this ratio.} 

%We will show that this ratio is free of ultraviolet (UV) divergences so that one can take the continuum limit of the lattice data smoothly. The subtraction of the soft factor is not necessary as well. Furthermore, perturbative matching is not needed at least at the one-loop level, which further simplifies the calculation.

%\section{TMD and quasi-TMD of linearly polarized gluons}

To start with, let us first review the gluon TMDs.
In QCD, the TMDs of the gluon in an unpolarized hadron with momentum $P$ are defined with the matrix element of the gluon field strength correlator~\cite{Mulders:2000sh}
\begin{align}
	&\int\frac{d\xi d^2 \vect b_T}{(2\pi)^3 P^+} e^{-i x \xi P^+ + i \vect{k}_T \cdot \vect{b}_T} \nonumber\\
	&\times \bigg\langle P\bigg|   {F_a^{+\mu}\left(\frac{\xi n+\boldsymbol{b}_{T}}{2}\right)W_{ab}^{-}F_b^{+\nu}\left(-\frac{\xi n+\boldsymbol{b}_{T}}{2}\right)}\bigg| P \bigg\rangle \nonumber \\
	=&-\frac{x}{2 }\bigg[g_T^{\mu \nu} f_1^g(x, \boldsymbol{k}_T^2)-\left(\frac{k_T^\mu k_T^\nu}{\boldsymbol{k}_T^2}+\frac12 g_T^{\mu \nu}  \right) \nonumber\\
	&~~~~\times h_1^{\perp g}(x,\boldsymbol{k}_T^2)\bigg]\, , \label{eq:lc:def:mom}
\end{align}
where $F_a^{\mu\nu}$ is the gluon field strength tensor in adjoint representation with $a$ being the color index, $\vect{k}_T$ is the transverse momentum of the gluon in the proton, $\xi$ and $\vect{b}_T$ are the displacements of fields along the $n$ and transverse directions, respectively. $f_1^g$ is the TMD for unpolarized gluons, while $h_1^{\perp g}$ is the TMD for linearly polarized gluons, $\boldsymbol{k}_T^2=-k_T^2$ and $g_T^{\mu\nu}=g^{\mu\nu}-n^{\mu}\bar n^{\nu}-n^{\nu}\bar n^{\mu}$ is the transverse metric, $n$ and $\bar n$ are two unit light-cone vectors. For any vector $a$, $n\cdot a=a^+$ and $\bar n\cdot a=a^-$. The Wilson line is generally process dependent; hereby we assume Wilson line $W^-_{ab}$ connects $\mp\frac{\xi n+\vect{b}_T}{2}$ via $-\infty$ along the $n$ direction.
 There are rapidity singularities in TMDs, and they can be renormalized by introducing the soft factors~\cite{Collins:2011zzd}. There are no model-independent calculation for $h_1^{\perp g}$ except its upper limit: $|h_1^{\perp g}(x,\vect{k}_T^2)|\leq f_1^g(x,\vect{k}_T^2)$~\cite{Mulders:2000sh}. 

In this work, we prefer to study the correlator in the $(x,\vect{b}_T)$-space, in which one can parameterize the correlator as
\begin{align}
	& \int\frac{d\xi}{2\pi P^+} e^{-i x \xi P^+} \nonumber\\
	&\times\bigg\langle P\bigg| F_a^{+\mu}\left(\frac{\xi n+\boldsymbol{b}_{T}}{2}\right)  {W_{ab}^-} F_b^{+\nu} \left(-\frac{\xi n+\boldsymbol{b}_{T}}{2}\right)\bigg| P\bigg\rangle \nonumber \\
	=&-\frac{x}{2 }\bigg[g_T^{\mu \nu} f_1^g(x, \vect{b}_T^2)-\left(\frac{b_T^\mu b_T^\nu}{\boldsymbol{b}_T^2}+\frac12 g_T^{\mu \nu}  \right) \nonumber\\
	&~~~~\times h_1^{\perp g}(x,\vect{b}_T^2)\bigg]\, .\label{eq:lc:def}
\end{align}
The TMDs in $(x,\vect{b}_T)$-space can be converted into TMDs in $(x,\vect{k}_T)$-space through Fourier-Bessel transforms.
The matrix element in Eq.~\eqref{eq:lc:def} contains correlations along the light-cone, which is hard to simulate on the Euclidean lattice.   

Instead, one can define a similar correlation matrix element but calculable on the lattice,
\begin{align}
	&\frac{P_3}{P_0^2}\int\frac{d\xi}{2\pi } e^{i x \xi P_3} \nonumber\\
	&\times \frac{\bigg\langle P\bigg| {E_{\perp a}^{i}\left(\frac{\xi n_z+\boldsymbol{b}_{T}}{2}\right)\widetilde{W}_{ab}^- E_{\perp b}^{j}\left(-\frac{\xi n_z+\boldsymbol{b}_{T}}{2}\right)}\bigg| P\bigg\rangle}{\sqrt{Z_E}}  \nonumber\\
	=&-\frac{x}{2 }\bigg[g_T^{i j} F_1^g(x, \boldsymbol{b}_T^2, P_3) -\left(\frac{b_T^i  b_T^j}{\boldsymbol{b}_T^2}+\frac12 g_T^{i j}  \right) \nonumber\\
	&~~~~\times H_1^{\perp g}(x,\vect{b}_T^2, P_3)\bigg]\, , \label{eq:ed:def}
\end{align}
where $n_z=(0,0,0,1)$ is the unit vector of the third Cartesian direction, $i, j= 1, 2$ denote the transverse components and $E_{\perp}^i= F^{0i} (i=1, 2)$ is the color electric field along the transverse directions.  The quasi-TMDs  $F_1^g$ and $H_1^{\perp g}$ are the Euclidean versions of $f_1^g$ and $h_1^{\perp g}$, respectively.  {Corresponding to the Wilson line structure in Eq.~\eqref{eq:lc:def}, the Wilson line $\widetilde{W}_{ab}^-$ is chosen as staple-shaped: from $-\frac{\xi}{2}n_z-\frac{\vect{b}_T}{2}$ to $(-L-\xi/2) n_z-\frac{\vect{b}_T}{2}$ along $-n_z$ direction, then from $(-L-\xi/2) n_z-\frac{\vect{b}_T}{2}$ to $(-L-\xi/2) n_z+\frac{\vect{b}_T}{2}$ along the transverse direction, then return to $\frac{\xi}{2}n_z+\frac{\vect{b}_T}{2}$ along $n_z$}.  
The UV singularities from the Wilson line self-interaction can be removed by a factor $\sqrt{Z_E}$  in the large $P_3$ (or small-$\xi$) limit. For the Wilson line structure described above, $Z_E$ can be chosen as a rectangular Euclidean Wilson-loop with length $2L$ and $|\vect{b}_T|$ (see Refs.~\cite{Ji:2019ewn,Zhu:2022bja}). 

 {On the lattice, one can adopt the  clover definition of field strength tensor in terms of plaquette which has been adopted in previous calculations, e.g., Refs.~\cite{Yang:2018bft,Shanahan:2018pib,ExtendedTwistedMass:2021rdx,Hackett:2023nkr,Good:2023ecp}.
	In our work only the color electric field $E$ is involved and can be expressed as $E_{i}=F_{0i}$.  It is related to the Euclidean operator $F_{4i}^{el}$ via $F_{4i}^{el}=-i F_{0i}$. 
	The Wilson line in adjoint representation can be expressed in terms of the Wilson lines in fundamental representation, through
	\begin{align}
		&F_a^{0i}\bigg(\frac{\xi n+\vect{b}_T}{2}\bigg) \widetilde{W}_{ab}^- F_b^{0j}\bigg(-\frac{\xi n+\vect{b}_T}{2}\bigg)\nonumber\\
		=&2 \operatorname{Tr}\bigg[F^{0i}\bigg(\frac{\xi n+\vect{b}_T}{2}\bigg) \widetilde{U}^- F^{0j}\bigg(-\frac{\xi n+\vect{b}_T}{2}\bigg) \widetilde{U}^{- \dag} \bigg]
	\end{align}
	where $\widetilde{U}^-$ is the Wilson line sharing the same path with $\widetilde{W}_{ab}^-$ but in the fundamental representation.}

In the infinite momentum frame, i.e., $P_3\to \infty$,  the operator in Eq.~\eqref{eq:ed:def} becomes the ``light-cone'' operator in Eq.~\eqref{eq:lc:def},  in which the third direction dependence becomes a light-cone dependence, and the color electric field $E^{i}$ becomes $F^{+i}$.  According to LaMET, the two matrix elements can be connected by perturbative matching, because $P_3\gg \Lambda_{\mathrm{QCD}}$ provides a hard scale.

Before moving on, we add some remarks on quasi-TMDs. First,  the choice of Euclidean correlation function is not unique. Any operator that approaches the operator in Eq.~\eqref{eq:lc:def} under large Lorentz boost can be used to define a quasidistribution. Second, there are UV divergences in Eq.~\eqref{eq:ed:def}, which may cause trouble for lattice calculations.  
%The gluon quasi-TMDs have already been introduced in Refs.~\cite{Schindler:2022eva} and \cite{Zhu:2022bja}. There are power-like UV divergences because of the space-like Wilson line, and such divergences have to be subtracted for practical calculations. This can be realized by subtracting the square root of a rectangular Wilson loop $Z_E$, see, e.g., Ref.~\cite{Ji:2019ewn}. But there are still UV divergences due to the interaction of the gluon field with the Wilson line.
There is no rapidity divergence in quasi-TMDs; however, a reduced soft factor should also be subtracted for a correct perturbative matching between TMDs and quasi-TMDs~\cite{Ji:2019sxk,Ji:2019ewn}.  
%Although the perturbative matching has been discussed in Ref.~\cite{Zhu:2022bja}, however, the renormalization is performed in $\overline{\mathrm{MS}}$ scheme, which is good for perturbative calculations but not convenient for lattice. 
 One may need some nonperturbative approaches to renormalize the UV singularities. In the case of quasi PDFs, DAs, and GPDs, several nonperturbative subtraction schemes have been employed, and have been applied to quark quasi-TMDs, such as RI/MOM scheme~\cite{Constantinou:2017sej,Alexandrou:2017huk,Shanahan:2019zcq}, ratio scheme~\cite{Radyushkin:2017cyf,Zhang:2022xuw}, hybrid scheme~\cite{Ji:2020brr}, etc. For the gluon TMD case, however, the large offshellness of the gluon in RI/MOM raises the risk of gauge invariance violation. The ratio scheme may work~\cite{Zhu:2022bja}, but calls for more nonperturbative inputs from the lattice.
%, and the theoretical foundation of ratio scheme is not established for gluon TMDs yet.

On the other hand,  we will not be troubled by the renormalization and soft factor subtraction issues, when we are studying the ratio of $H_1^{\perp g}$ and $F_1^{g}$: $R\equiv H_1^{\perp g}/F_1^{g}$, as we will discuss below. 
Various ratios have been constructed on the lattice for quark TMDs~\cite{Musch:2011er,Engelhardt:2015xja,Ebert:2020gxr,Vladimirov:2020ofp} before.
Our ratio $R(x, \vect{b}_T^2, P_3)$ can be expressed in terms of operator matrix elements as 
\begin{widetext}
\begin{align}
	&\frac12+\frac14 R(x, \boldsymbol{b}_{T}^2, P_3 )=\frac{\int\frac{d\xi}{2\pi} e^{i x\xi P_3}\langle P| \boldsymbol{b}_T\cdot \boldsymbol{E}_{\perp a}(\frac{\xi}{2} n_z+\frac{\boldsymbol{b}_T}{2}) {\widetilde{W}_{ab}^-} \boldsymbol{b}_T\cdot \boldsymbol{E}_{\perp b}(-\frac{\xi}{2}n_z-\frac{\boldsymbol{b}_T}{2}) |P\rangle}{\int\frac{d\xi}{2\pi} e^{i x \xi P_3}\boldsymbol{b}_T^2\langle P| \boldsymbol{E}_{\perp a} (\frac{\xi}{2} n_z+\frac{\boldsymbol{b}_T}{2})\cdot  {\widetilde{W}_{ab}^-}\boldsymbol{E}_{\perp b}(-\frac{\xi}{2}n_z-\frac{\boldsymbol{b}_T}{2})|P\rangle }\, . \label{eq:R}
\end{align}
\end{widetext}
Its light-cone partner, $h_1^{\perp g}/f_1^g$, is the relative strength of the linearly polarized gluons over the unpolarized gluons, which is a reflection of the degree of polarization.
In the infinite momentum limit, one can expect that
$R(x, \boldsymbol{b}_{T}^2, P_3 )\to h_1^{\perp g}(x, \vect{b}_{T}^2)/ f_1^g(x, \vect{b}_T^2)  $.
According to Eq.~\eqref{eq:R}, $R$ is a ratio of Euclidean correlation functions and there is no time-dependence, thus it can be simulated on the lattice.

%\section{Cancellation of UV divergences}

For a practical calculation on the lattice, one has to renormalize the quantities properly, because the UV singularities prevent taking the continuum limit of the lattice data.
The renormalization of gluonic Wilson line operators has been studied a long time ago~\cite{Dorn:1986dt}, and recently has been revisited in the context of quasi-PDF by using the auxiliary field formalism~\cite{Zhang:2018diq} and also the pseudo-PDF approach~\cite{Balitsky:2019krf,Balitsky:2021qsr,Balitsky:2021cwr}. There is no essential difference between the ``staple-shaped''  operators here and the ``straight line'' operators in quasi-PDF on the renormalization of UV singularities. 

There are three sources of UV singularities: the self-energy of gluon, the self-interaction of the Wilson line, and the interaction between the Wilson-line and the field operator located at the endpoint. The gluon self-energy is canceled in the ratio.
The UV singularities from the self-interaction of the Wilson line are multiplicatively renormalized, even if there are cusps in the Wilson line. For the Wilson line described in the last paragraph, in the large $P_3$ limit, the UV singularities from the Wilson line self interaction can be removed by the factor $\sqrt{Z_E(2L, \vect{b}_{\perp})}$, where $Z_E$ is a rectangular Euclidean Wilson-loop with length $2L$ and $|\vect{b}_T|$ (see Refs.~\cite{Ji:2019ewn,Zhu:2022bja}). This factor cancels between the numerator and denominator.
 The interaction between the Wilson line and the field located at the endpoint may lead to operator mixing; however, the operator 
is multiplicatively renormalizable if the field operator located at the endpoint is $F^{0i}$, $F^{3i}$ or $F^{3\mu}$, where $i=1,2$ and $\mu=0,1,2$~\cite{Zhang:2018diq}.  
In addition, the renormalization factor is independent of the location of the operator, which means that the Fourier transform does not modify the multiplicative renormalizability. 
In Eq.~\eqref{eq:R}, the UV divergences in the denominator and numerator are multiplicative, and the renormalization factors are equal 
because the operators in both the denominator and numerator are of the $F^{0i}F^{0j}$ ($i,j=1,2$) type and the Wilson line structures are the same. For the above reasons,  the ratio Eq.~\eqref{eq:R} is UV finite, because
all UV singularities, including cusp and pinched pole singularities, as well as the endpoint UV singularities,  are canceled in the ratio. So,  the continuum limit of $R(x,\vect{b}_T,P_3)$ can be approached without a renormalization procedure on the lattice.
% We may go even further, that the ratio may be independent of, or at least not sensitive to the shape of the Wilson line. 

%\section{Perturbative matching}

In LaMET, the Euclidean and light-cone quantities are linked by a  matching relation, while the matching coefficient can be calculated in perturbation theory because it is associated with a hard scale $P_3$. It has been shown that the TMD matching in LaMET has the type of multiplication instead of a convolution. This is  confirmed in the case of gluon TMD~\cite{Schindler:2022eva,Zhu:2022bja}, where the matching for gluon TMD was derived as 
\begin{align}
	&F_1^g\left(x, \vect{b}_{T}^2 , \mu, \zeta_{z}\right) S_{r}^{\frac{1}{2}}\left(\vect{b}_{T}^2, \mu\right)  \nonumber\\
	=& H\left(\frac{\zeta_{z}}{\mu^{2}}\right) e^{\ln \frac{\zeta z}{\zeta} K\left(\vect{b}_{T}^2, \mu\right)} f_1^g\left(x, \vect{b}_{T}^2, \mu, \zeta\right)\, , \label{eq:matchingformula}
\end{align}
where $S_r$ is the reduced soft factor, $K$ is the Collins-Soper kernel and $H$ is the hard function, $\zeta_z=(2 x P_3)^2$ and $\zeta$ is the Collins-Soper scale.  The matching relation for $H_1^{\perp}$ is the same but the hard function may differ.  
Generally, we have the matching relation
\begin{align}
	R(x, \vect{b}_T^2, P_3)=\frac{H_h\left(\frac{\zeta_z}{\mu^2}\right)}{H_f\left(\frac{\zeta_z}{\mu^2}\right)} \frac{h_1^{\perp g} (x,\vect{b}_T^2 ,\mu, \zeta)}{f_1^{g} (x,\vect{b}_T^2 ,\mu, \zeta)}\, , \label{eq:match}
\end{align}
where $H_h$ and $H_f$ are matching coefficients for $f_1^g$ and $h_1^{\perp g}$, respectively.  The $S_r$ and $K$ terms cancel in the matching formula. Thus we do not need to worry about these quantities, which makes the evaluation simpler. 

\begin{figure}
	\centering
	\includegraphics[width=0.9\linewidth]{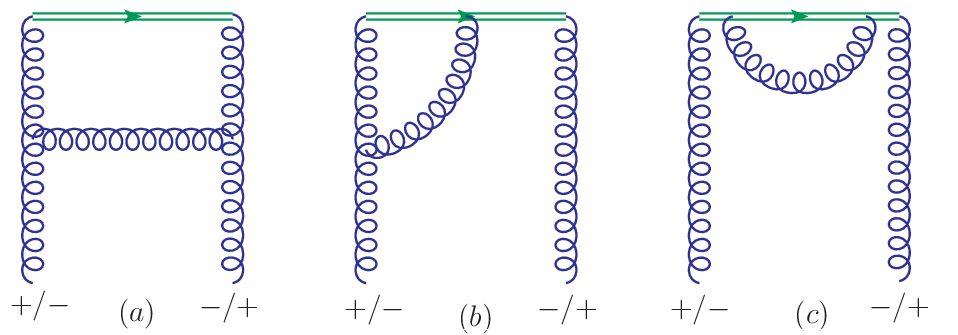}
	\caption{The typical Feynman diagrams for one-loop correction of the operators in Eq.~\eqref{eq:lc:def} and Eq.~\eqref{eq:ed:def} in Feynman gauge. The other Feynman diagrams are not shown.}
	\label{fig:diagram}
\end{figure}

The matching for the denominator in Eq.~\eqref{eq:R} has already been studied and $H_f$ has been calculated at the one-loop level. Now we will derive the matching relation for the numerator.  To perform the matching calculation in perturbation theory, one can replace the hadron state with a parton state because the hard function is independent of external states.
In previous works, the external states are always chosen as unpolarized gluons. It was shown in~\cite{Ma:2012hh} that $h_{1}^{\perp g}$ in unpolarized gluon target is $x h_1^{\perp g} (x, \vect{k}_T^2)=  2\alpha_s C_A(1-x)/(\pi^2 \vect{k}_{T}^2)+\mathcal{O}(\alpha_s^2)$, in which the nonzero result starts at one-loop level, and only box diagram (see Fig.~\ref{fig:diagram}(a)) has nonzero contribution. So, if the external gluon is unpolarized, one can only work out the matching coefficient by calculating at least two-loop diagrams, which will be a rather tough task.

Instead, we assume that the external gluons are emitted from an unpolarized hadron and they are polarized, then extract  $h_1^{\perp g}$ and $H_1^{\perp g}$ by calculating the helicity-flip matrix element, i.e.,  $\langle p,-| \cdots | p, +\rangle -\langle p,+| \cdots | p, -\rangle $, where $+/-$ denotes the gluon helicity $+1$ or $-1$. 
The amplitude for general gluon helicities can be expressed as $\mathcal {M}_{ij}\epsilon_{1}^i\epsilon_{2}^{*j}$, then the hecility-flip contribution we needed is $\mathcal {M}_{ij}(\epsilon_{+}^i\epsilon_{-}^{*j} -\epsilon_{-}^j\epsilon_{+}^{*i} )$.
 One can replace the gluon density matrix $\epsilon_{1}^i\epsilon_{2}^{*j}$ with $\frac12 (b_T^i b_T^j/\vect{b}_T^2+\frac12 g_T^{ij})$ to simplify the calculation.
The tree-level result is no longer zero but $\delta(1-x)$. 
At one-loop, one can perform a one-loop calculation in dimensional regularization, in which the dimensions of spacetime are $d=4-2\epsilon$.
The decomposition of correlator in Eqs.~\eqref{eq:lc:def}\eqref{eq:ed:def} in $d$-dimensions then becomes
$$
-\frac{1}{d-2} \bigg[g_T^{ij} f_1^g -\left(\frac{b_T^i  b_T^j}{\boldsymbol{b}_T^2}+\frac{1}{d-2} g_T^{i j}  \right) h_1^{\perp g}\bigg]\, .
$$
In Fig.~\ref{fig:diagram}, we list three typical Feynman diagrams in the Feynman gauge at the one-loop level. Here we adopt the procedure in Ref.~\cite{Zhu:2022bja}. 
All the Feynman diagrams are categorized into three classes: (a): No Wilson line interaction; (b): Involving gluon-Wilson line interactions and (c): Wilson line self-interaction. For (a), we find that the TMD and quasi-TMD have the same results, and thus have no contribution to the matching coefficient;  (c) has no contribution if the reduced soft factor is subtracted. Because we are discussing the ratio, we do not need to consider the soft factor at hand because they are canceled in the ratio. (b) involves rapidity singularities and contributes to the matching.  
After some tedious but straightforward calculation, we find that the total result for both $H_1^{\perp g}$ at large $P_3$ and $h_1^{\perp g}$ have the structure
\begin{align}
	-\frac{\alpha_s }{2\pi}C_A & \theta(0<x\leq 1)\bigg[\frac{2x}{(1-x)_+}+\frac{\beta_0}{2C_A}\delta(1-x)\bigg]\nonumber\\
	&\times \left(\frac{1}{\epsilon_{\mathrm{IR}}}+\ln\frac{\mu^2 \vect{b}_T^2 e^{2\gamma_E}}{4}\right) +\delta(1-x) \mathcal{C}\, ,
\end{align}
where ``$+$'' denotes the plus distribution, $\beta_0=\frac{11}{3}C_A-\frac43 T_F n_f$. Note that the above expression is defined in the support $[0,1]$. The values of constant $\mathcal{C}$ are
\begin{subequations}
	\begin{align}
	&	\mathcal{C}_H=\frac{\alpha_s}{2\pi}C_A \bigg[\bigg(\frac{1}{\epsilon_{\mathrm{UV}}}+\ln\frac{\mu^2 \vect{b}_T^2  e^{2\gamma_E}}{4}\bigg)\bigg(
	\frac{\beta_0}{2C_A}-1\bigg) \nonumber\\
		&~~~~-\frac{1}{2}\ln^2 (p_3^2 \vect{b}_T^2  e^{2\gamma_E})  +2\ln  (p_3^2 \vect{b}_T^2 e^{2\gamma_E})   -4\bigg]\, ,\\
		&\mathcal{C}_h=\frac{\alpha_s}{2\pi}C_{A} \bigg[\frac{1}{\epsilon_{\mathrm{UV}}^2}+ \left(\frac{1}{\epsilon_{\mathrm{UV}}}+\ln\frac{\mu^2 \vect{b}_T^2  e^{2\gamma_E}}{4} \right) \nonumber\\
		&~~~~\times \left( \frac{\beta_{0}}{2 C_A}+ \ln \frac{\mu^{2}}{\zeta} \right) -\frac{1}{2}\ln^2\frac{\mu^2 \vect{b}_T^2  e^{2\gamma_E}}{4}-\frac{\pi^{2}}{12}\bigg] 
	\end{align}
\end{subequations}
for quasi-TMD and  TMD, respectively. The result for TMD here is subtracted by the soft factor; however, subtracting the soft factor or not does not affect the matching of the ratio. Because the IR structure of the normal and quasi-TMDs are the same, their differences are only related to UV and  the matching coefficients read
\begin{align}
	&H_h \left(\frac{\zeta_{z}}{\mu^{2}}\right) =H_f \left(\frac{\zeta_{z}}{\mu^{2}}\right)\nonumber\\
	=&1
	+
	\frac{\alpha_s }{2\pi} C_A\left( -\frac{1}{2}\ln^2 \frac{\zeta_z}{\mu^{2}}+2\ln \frac{\zeta_z}{\mu^{2}}+\frac{\pi^{2}}{12}-4\right)\nonumber\\
	&~+\mathcal{O}(\alpha_s^2)\, , \label{eq:coef}
\end{align}
where $\zeta_z=(2 x P_3)^2$. The matching coefficients for $f_1^g(x,\vect{b}_T^2)$ and $h_1^{\perp g}(x,\vect{b}_T^2)$ are equal at one-loop accuracy.   { It was shown in a previous work~\cite{Zhu:2022bja} that the matching coefficient for the helicity gluon TMD is also equal to the one in Eq.~\eqref{eq:coef}, and it is likely that the matching coefficient might be equal for all of the right gluon TMDs at one-loop.}
  {It is different from the one-dimensional distributions, for example, the matching kernel for  unpolarized and helicity gluon PDFs are not equal.  Additionally, it is not clear whether Eq.~\eqref{eq:coef} holds at all orders of $\alpha_s$. The differences between the matching of TMDs and one-dimensional PDFs require further explorations.}
 
According to Eq.~\eqref{eq:match} and Eq.~\eqref{eq:coef}, one can conclude that $H_h/H_f=1+\mathcal{O}(\alpha_s^2)$, and
\begin{align}
	&h_1^{\perp g} (x,\vect{b}_T^2 ,\mu, \zeta) / f_1^g (x,\vect{b}_T^2 ,\mu, \zeta)\nonumber\\
	&~~\simeq R(x, \vect{b}_T^2, P_3)+\mathcal{O}(\alpha_s^2) +\mathcal{O}\left( \frac{\Lambda_{\mathrm{QCD}}}{x P_3}\right)\, . \label{eq:ratio:fac}
\end{align}  
%This indicates that the lattice results for $R(\xi, \vect{b}_T^2, P_3)$ is a very nice approximation of the degree of polarization because the perturbative matching effect is suppressed by $\alpha_s^2$. 

We note that the matching coefficient in Eq.~\eqref{eq:coef} is derived in the $\overline{\mathrm{MS}}$ scheme. 
%In a practical lattice calculation, the commonly used schemes include the ratio scheme, RI/MOM, hybrid scheme, etc.  
In our proposal, we do not need lattice renormalization schemes because the ratio is UV finite, and although different results for hard functions $H_h$ and $H_f$ may be derived in different schemes, their ratio should be equal.

The determination of ratio $R$ on the lattice is helpful in phenomenology at the hadron colliders. We take the production of scalar (or pseudoscalar) boson $H$ at low transverse momentum as an example.  By converting the factorization formula in $(x,\vect{p}_{T})$-space in Ref.~\cite{Boer:2011kf} to $(x,\vect{b}_T)$ space, the differential cross section at low $|\vect{q}_T|$ can be expressed in terms of $f_1^g$ and $R$.  {To get rid of the $f_1^g$ part, one can further introduce the ratio with the differential cross section of $J/\Psi+\gamma$~\cite{denDunnen:2014kjo}:}
\begin{align}
	&\frac{\int\frac{d^2 \vect{q}_T}{(2\pi)^2} e^{i \vect{q}_T\cdot \vect{b}_T} \frac{d \sigma(A+B \rightarrow H+X)}{d x d y d^2 \boldsymbol{q}_T}}{\int\frac{d^2 \vect{q}_T}{(2\pi)^2} e^{i \vect{q}_T\cdot \vect{b}_T} \frac{d \sigma(A+B \rightarrow J/\Psi+\gamma+X)}{d x d y d^2 \boldsymbol{q}_T}} \nonumber\\
	&\propto 1 \pm \frac{1}{4} R\left(x, \boldsymbol{b}_T^2, P_3\right) R\left(y, \boldsymbol{b}_T^2, P_3\right),
\end{align}	
%\begin{align}
%	&\frac{d \sigma(A+B\to H+X)}{dx dy d^2 \vect{q}_T}  
%	\propto \int d^2 \vect{b}_T e^{-i \vect{b}_T\cdot \vect{q}_T} \nonumber\\
%	& f_1^g (x, \vect{b}_T)f_1^g (y,\vect{b}_T)\bigg[1\pm \frac14 R(x, \vect{b}_T^2, P_3)R(y, \vect{b}_T^2, P_3)   \bigg]\, 
%\end{align}
where in the denominator $\vect{q}_T$ is the transverse momentum of the $J/\Psi\gamma$ pair, and differential cross sections should be measured at low $\vect{q}_T$. $+/-$ corresponds to the scalar and pseudo scalar, respectively. Thus the effect of linearly polarized gluon on the Higgs boson production could be determined with lattice QCD calculations.  With similar discussions in~\cite{Boer:2011kf}, our ratio can also be used to determine the parity of the Higgs boson.
%\section{Conclusion}

To summarize, we have explored the feasibility of calculating the TMD of linearly polarized gluons in an unpolarized hadron on the lattice, in the framework of large momentum effective theory. We propose to calculate the ratio of linearly polarized gluon TMD over the unpolarized gluon TMD, which characterizes the degree of gluon polarization. We define a Euclidean version of this ratio, which is UV finite. Therefore, no renormalization and soft factor subtraction are necessary. Furthermore, we evaluate the perturbative matching that connects the ratio and its light-cone partner and find that the perturbative matching coefficient is zero at one-loop.  Thus the ratio discussed in this work is a good approximation of the ratio of $h_1^{\perp g}$ and $f_1^g$. 
Future lattice simulations will shed light on the distribution of linearly polarized gluons in a hadron, and could provide useful information for phenomenology at the hadron colliders.

%\section*{Acknowledgements}
I thank Yao Ji, Jian-Hui Zhang, and Ruilin Zhu for useful discussions and collaboration on~\cite{Zhu:2022bja} which inspires the present work.
This work is supported by Jefferson Science Associates,
 LLC under  U.S. DOE Contract \#DE-AC05-06OR23177
 and by U.S. DOE Grant \#DE-FG02-97ER41028.

\appendix

\bibliographystyle{apsrev}
\bibliography{ref}

\end{document}